\documentclass[twocolumn,showpacs,showkeys,preprintnumbers]{revtex4}

\def\be{\begin{equation}}
\def\ee{\end{equation}}
\def\beq{\begin{eqnarray}}
\def\eeq{\end{eqnarray}}
\def\n{\nonumber}
\def\bay{\begin{array}}
\def\eay{\end{array}}

\begin{document}

\preprint{CIRI/02-swkg02}
\title{Radially homothetic spacetime is of Petrov-type D}

\author{Sanjay M. Wagh
$^1$ and Keshlan S. Govinder$^2$}

\affiliation{$^{1}$Central India Research Institute, Post Box 606,
Laxminagar, Nagpur 440 022, India\\
E-mail:cirinag@nagpur.dot.net.in \\
$^2$School of Mathematical \& Statistical Sciences, University of
Natal, Durban 4041, South Africa.
\\E-mail: govinder@nu.ac.za }

\date{September 9, 2002}
\begin{abstract}
It is well-known \cite{mtbh} that {\em all\/} black hole solutions
of General Relativity are of Petrov-type D. It may thus be
expected that the spacetime of {\em physically realizable\/}
spherical gravitational collapse is also of Petrov-type D. We show
that a radially homothetic spacetime, {\em ie}, a spherically
symmetric spacetime with hyper-surface orthogonal, radial,
homothetic Killing vector, is of Petrov-type D. As has been argued
in \cite{prl1}, it is a spacetime of {\em physically realizable}
spherical collapse.
\end{abstract}

\pacs{04.20.-q, 04.20.Cv}%
\keywords{Gravitational collapse  - spherical symmetry - radial
homothety - Petrov-type D}%
\maketitle

\newpage

\section{Introduction}
In General Relativity, all black hole solutions are of Petrov-type
D \cite{mtbh}. Then, it may be expected that the spacetime of
``physically realizable" spherical collapse of matter is of
Petrov-type D.

In a recent work \cite{cqg1}, we obtained a spherically symmetric
spacetime by considering a metric separable in co-moving
coordinates and by imposing a relation of pressure $p$ and density
$\rho$ of the barotropic form $p=\alpha\rho$ where $\alpha$ is a
constant. Such a relation determined only the temporal metric
functions of that spacetime.

Therefore, although the temporal metric functions for this
spacetime were determined in \cite{cqg1} using the above
barotropic equation of state, same metric functions are
determinable from {\em any\/} relation of pressure and density.

This spacetime admits a hyper-surface orthogonal, radial,
Homothetic Killing Vector (HKV). (See later.) Hence, it will be
called a {\em radially homothetic spacetime}. Since this spacetime
describes appropriate ``physical" stages of evolution of spherical
matter, we have argued in \cite{prl1} that it is the spacetime of
{\em physically realizable\/} spherical collapse of matter. As we
show in this letter, it is a Petrov-type D spacetime.

\section{Spacetime metric} \label{spacetime}
In general, a HKV captures \cite{carrcoley} the notion of the
scale-invariance of the spacetime. A homothetic spacetime then
admits an appropriate homothetic Killing vector ${\bf X}$
satisfying \be {\cal L}_{\bf X} g_{ab}\;=\;2\,\Phi\,g_{ab}
\label{ckveq1} \ee where $\Phi$ is an arbitrary constant.

If, in terms of the chosen coordinates, ${\bf X}$ has component
only in the direction of one coordinate, the Einstein field
equations separate for that coordinate, generating also an
arbitrary function of that coordinate. This is the broadest (Lie)
sense of the scale-invariance leading not only to the reduction of
the field equations as partial differential equations to ordinary
differential equations but leading also to their separation.

A spherically symmetric spacetime has only one spatial scale
associated with it - the radial distance scale. Therefore, for a
{\em radially homothetic spacetime}, the metric admits one
arbitrary function of the radial coordinate. We then obtain for a
radially homothetic spacetime arbitrary radial characteristics for
matter. That is, due to the radial scale-invariance of the
spherical spacetime, matter has {\em arbitrary\/} radial
properties in a radially homothetic spacetime.

A general spherically symmetric spacetime admits, in co-moving
coordinates, a metric of the form \be ds^2 = -e^{2\nu(r,t)}dt^2
+e^{2\lambda(r,t)}dr^2+Y^2(r,t)d\Omega^2 \label{genssmet} \ee
where $d\Omega^2 = \left[ \,d\theta^2\,+\,\sin^2{\theta}\,
d\phi^2\,\right]$.

In what follows, we therefore demand that a spherically symmetric
spacetime of (\ref{genssmet}) admits a spacelike HKV of the form
\be X^a\;=\;(0,X^1(r,t),0,0) \label{hkvradial1} \ee Then, the
expression (\ref{ckveq1}) reduces to the system of four equations:
\beq
X^1\, \nu_{,\,r}&=& \Phi \label{phinu} \\
X^1_{,\,t} &=& 0 \label{fcr}\\
\left( \frac{Y_{,\,r}}{Y}\,-\,
\nu_{,\,r} \right)\, X^1 &=& 0 \\
\left( \lambda_{,\,r}\,-\, \nu_{,\,r} \right)\, X^1\;+\;
X^1_{,\,r} &=&0 \eeq where a comma denotes a derivative and $\Phi$
is a constant. (See \cite{moopanar}, for the comprehensive and
detailed discussion of the general conformal geometry of the
metric (\ref{genssmet}).)

Solving the above system of equations for $X^1$, $\nu$, $\lambda$
and $Y$, we obtain \beq X^1 &=& F(r) \label{fofr}
\label{feqn} \\
Y &=& \tilde{g}(t)\,\exp\left(\int\frac{\Phi}{F(r)}d r\right)
\\ \lambda &=& \int\frac{\Phi}{F(r)}d r \;-\;\log F(r) \;+\; \tilde{h}(t) \\
\nu &=& \int\frac{\Phi}{F(r)}d r + \tilde{k}(t)  \eeq where $F(r)$
is an arbitrary function of the co-moving radial coordinate $r$.
We emphasize here that the HKV (\ref{hkvradial1}) is expected, on
the basis of Lie's theory of differential equations, to generate
an arbitrary function of the radial coordinate $r$ for the
spacetime. This is seen to be the case.

Then, the spacetime metric becomes {\it separable\/} in co-moving
coordinates and is given by \beq ds^2&=& k^2(t)
\exp{\left(2\int\frac{\Phi}{F(r)}d
r\right)}\left[-\,dt^2 \phantom{\frac{h^2(t)}{F^2(r)}} \right. \n \\
 &&\left. \qquad \quad +\frac{h^2(t)}{F^2(r)}\,dr^2 +\;g^2(t)\,d\Omega^2 \right]
\label{ssmetfinal0} \eeq

Now, writing $F(r)=y(r)\Phi/ y'$, we obtain \be
ds^2=-y^2dt^2+\gamma^2(y')^2B^2dr^2+y^2Y^2d\Omega^2
\label{ssmetfinal}\ee with a prime indicating a derivative with
respect to $r$, $B\equiv B(t)$, $Y\equiv Y(t)$ and $\gamma=1/\Phi$
is a constant. (Temporal function in $g_{tt}$ is absorbed by
suitable redefinition of the time coordinate.)

Therefore, the imposition of HKV (\ref{hkvradial1}) {\it
uniquely\/} determines, in co-moving coordinates, the spherically
symmetric metric to (\ref{ssmetfinal}) \cite{cqg1}.

Now, the radial scale-invariance of a spherical spacetime
identifies the spacetime metric uniquely. To see this, let us use
a gaussian radial coordinate and a area radial coordinate.

In a gaussian coordinate system, the most general spherical metric
is \cite{synge} \be ds^2 =
-\,e^{2\nu(\bar{r},\tau)}d\tau^2+d\bar{r}^2+\bar{r}^2f^2(\bar{r},\tau)d\Omega^2\ee
The radial scale-invariance then demands the existence of a HKV of
the form $(0,X^1(\bar{r},\tau),0,0)$. Then, (\ref{ckveq1}) implies
\beq \nu_{,\,\bar{r}}\,X^1&=&\Phi \\ X^1_{,\,\tau} &=&0
\\X^1_{,\,\bar{r}}&=&\Phi \label{coneq} \\\left( \frac{1}{\bar{r}}+\frac{f_{,\,\bar{r}}}
{f}\right) X^1 &=& \Phi \eeq The spacetime metric of a radially
homothetic, spherical spacetime in a gaussian coordinate system is
then uniquely obtained as \beq ds^2 &=& -\,(\Phi
\bar{r}+K_1)^2k^2(\tau)d\tau^2 + d\bar{r}^2 \n \\ && \qquad\qquad
+\,(\Phi \bar{r}+K_1)^2h^2(\tau)d\Omega^2 \label{ssmetgauss} \eeq
where $K_1$ is a constant of integration from (\ref{coneq}).

Further, for the curvature or the area coordinates, the general
spherical metric is \cite{synge} \be ds^2=-\,e^{2\nu(\tilde{r},
\bar{t})}d\bar{t}^2 + e^{2\lambda(\tilde{r}, \bar{t})}d\tilde{r}^2
+ \tilde{r}^2d\Omega^2 \ee The radial scale-invariance then
requires us to impose the HKV $(0,X^1(\tilde{r},\bar{t}),0,0)$.
Then, (\ref{ckveq1}) becomes \beq \nu_{,\,\tilde{r}}\,X^1&=&\Phi \\
X^1_{,\,\bar{t}} &=&0
\\X^1_{,\,\tilde{r}}+\lambda_{,\,\tilde{r}}&=&\Phi \\\frac{X^1}{\tilde{r}}
&=& \Phi \eeq We then, uniquely, obtain the metric \be ds^2= -
\tilde{r}^2h^2(\bar{t})d\bar{t}^2 + g^2(\bar{t})d\tilde{r}^2 +
\tilde{r}^2d\Omega^2 \label{ssmetarea} \ee

As can be easily verified, the three forms above, namely,
(\ref{ssmetfinal}), (\ref{ssmetgauss}) and (\ref{ssmetarea}), are
diffeomorphically equivalent to each other.

We select the metric form (\ref{ssmetfinal}) in co-moving
coordinates for our present study since it explicitly displays the
radial scale-invariance of the spacetime under consideration here.

Therefore, a radially homothetic, spherical spacetime admits a
spacelike HKV of the form \be X^a\;=\;(0,\frac{y}{\gamma y'},0,0)
\label{hkvradial} \ee and the spacetime metric is then given by
(\ref{ssmetfinal}).

Now, the spacetime of (\ref{ssmetfinal}) is required, by
definition, to be locally flat at all of its points including the
center.

The condition for the center to possess a locally flat
neighborhood is \be {y'|}_{r\,\sim\,0}\;\approx\;1/\gamma
\label{conrzero} \ee This condition must be imposed on any $y(r)$.
With this condition, (\ref{conrzero}), the HKV of metric
(\ref{ssmetfinal}) is, at the center,
$y|_{r=0}\,\partial/\partial_r$.

Now,  $y(r)$ is the ``area radius" in (\ref{ssmetfinal}). When
$y|_{r=0} \neq 0$, the orbits of the rotation group $SO(3)$ do not
shrink to zero radius at the center for (\ref{ssmetfinal}).
Consequently, the center is not regular for (\ref{ssmetfinal})
when $y|_{r=0} \neq 0$ although the curvature invariants remain
finite at the center.

Also, when $y|_{r=0} = 0$, the center is regular for the spacetime
of (\ref{ssmetfinal}). But, the curvature invariants blow up at
the center, then.

It is well-known \cite{mcintosh} that the center and the initial
data for matter, both, are not {\em simultaneously\/} regular for
a spherical spacetime with hyper-surface orthogonal HKV.
Therefore, the spacetime of (\ref{ssmetfinal}) does not possess a
{\em regular\/} center and {\em regular\/} matter data,
simultaneously.

However, the lack of regularity of the center of
(\ref{ssmetfinal}) for non-singular matter data is understandable
\cite{prl1} since the orbits of the rotation group do not shrink
to zero radius for every observer. It is a relative conception and
the co-moving observer, it being a ``cosmological" observer, of
(\ref{ssmetfinal}) is not expected to observe the orbits of the
rotation group shrink to zero radius.

Further, the spacetime of a spherical body must possess
non-vanishing central value for mass in it. In Newtonian gravity,
this is the theorem: ``The gravitational force on a body that lies
outside a closed spherical shell of matter is the same as it would
be if all the shell's matter were concentrated into a point at its
center."

Now, the mass function for the spacetime of (\ref{ssmetfinal}) can
be defined as \be m(r,t)
\;=\;\frac{yY}{2}\,\left(\,1\,-\,\frac{Y^2}{\gamma^2B^2}\,+\,\dot{Y}^2\,\right)
\label{mass} \ee where $m(r,t)$ denotes the total mass in the
spacetime, ie, mass of matter together with the ``effective''
contribution due to the flux of radiation or heat in the
spacetime.

Then, all points, including the center, of the ``initial"
spacelike hyper-surface, at $t=0$, evolve along the respective
timelike trajectories and, with all points of the hyper-surface,
non-vanishing mass exists also at the center. This is consistent
with the Newtonian theorem mentioned earlier.

\section{Singularities and degeneracies of the metric (\ref{ssmetfinal})}

The Ricci scalar for (\ref{ssmetfinal}) is: \beq {\cal R} &=&
\frac{4\dot{Y}\dot{B}}{y^2YB}+\frac{2\ddot{B}}{y^2B}
-\frac{6}{y^2\gamma^2B^2}\n \\
&&\qquad\qquad  +\frac{2}{y^2Y^2} +\frac{2\dot{Y}^2}{y^2Y^2}
+\frac{4\ddot{Y}}{y^2Y} \label{ricciscalar} \eeq

The Einstein tensor for (\ref{ssmetfinal}) is: \beq G_{tt}&=&
\frac{1}{Y^2}-\frac{1}{\gamma^2B^2} + \frac{\dot{Y}^2}{Y^2} +
2\frac{\dot{B}\dot{Y}}{BY}
\\ G_{rr}&=& \frac{\gamma^2B^2y'^2}{y^2} \left[-\,2\frac{\ddot{Y}}{Y}
-\frac{\dot{Y}^2}{Y} \right. \n \\ && \qquad \qquad \qquad \left.
+\frac{3}{\gamma^2B^2} - \frac{1}{Y^2}\right]
\\G_{\theta\theta}&=&-\,Y\,\ddot{Y}-Y^2\frac{\ddot{B}}{B}
- Y\,\frac{\dot{Y}\dot{B}}{B}+\frac{Y^2}{\gamma^2B^2}
\\G_{\phi\phi}&=& \sin^2{\theta}\,G_{\theta\theta} \\
G_{tr}&=&2\frac{\dot{B}y'}{By}  \label{gtr} \eeq where an overhead
dot denotes a time-derivative.

Now, for the co-moving observer with four-velocity $ {\bf
U}\,=\,\frac{1}{y}\;\frac{\partial}{\partial t}$, the {\em
radial\/} velocity of the fluid is $ V^r =\dot{Y}$. The co-moving
observer is accelerating for (\ref{ssmetfinal}) since
$\dot{U}_a={U_a}_{;\,b}U^b$ is, in general, non-vanishing for
$y'\neq 0$. The expansion is $ \Theta =
\frac{1}{y}\,\left(\,\frac{\dot{B}}{B} \;+\;
2\,\frac{\dot{Y}}{Y}\,\right)$. Further, $\dot{B}$ is related to
the flux of radiation in the co-moving frame.

Clearly, we may use the function $y(r)$ in (\ref{ssmetfinal}) as a
new radial coordinate - the area coordinate - as long as $y' \neq
0$. However, the situation of $y'=0$ represents a coordinate
singularity that is similar to, for example, the one on the
surface of a unit sphere where the analogue of $y$ is
$\sin{\theta}$ \cite{synge}. The curvature invariants do not blow
up at locations for which $y'=0$.

Genuine curvature singularities exist for (\ref{ssmetfinal}) when
either $y(r)\,=\,0$ for some $r$ or when the temporal functions
vanish for some $t=t_s$.

There are, therefore, two types of curvature singularities of the
spacetime of (\ref{ssmetfinal}), namely, the first type for
$B(t_s)=0$ and, the second type for $y(r)=0$ for some $r$.

Note that the ``physical'' radial distance corresponding to the
``coordinate'' radial distance $\delta r$ is \be \ell\;=\;\gamma
(y') B \delta r \ee Then, collapsing matter forms the spacetime
singularity in (\ref{ssmetfinal}) when $B(t)\,=\,0$ is reached for
it at some $t\,=\,t_s$. Therefore, the singularity of first type
is a singular hyper-surface for (\ref{ssmetfinal}).

The singularity of the second type is a singular sphere of
coordinate radius $r$. The singular sphere reduces to a singular
point for $r=0$ that is the center of symmetry. For $y(r)=0$ for
some range of $r$, there is a singular thick shell. Singularities
of the second type constitute a part of the initial data, singular
data.

The metric (\ref{ssmetfinal}) has evident degeneracies when
$y(r)={\rm constant}$ for some range of the co-moving radial
coordinate $r$ or globally. Further, the metric (\ref{ssmetfinal})
is also degenerate for $y(r)= \infty$ either on a degenerate
sphere of coordinate radius $r$, for some ``thick shell" or
globally. The degeneracy $y(r)=\infty$ is equivalent to vacuum.
For $y(r)=constant$, the degeneracy corresponds to uniform
density. For $y(r)=0$, the degeneracy is also an infinite density
singularity.

{\em In what follows, we shall assume, unless stated explicitly,
that there are no singular initial-data and that there are no
degenerate situations for the metric (\ref{ssmetfinal}).}

\section{Spacetime of the metric (\ref{ssmetfinal}) is of Petrov-type D}
In what follows, we use $\sigma$ to denote the null quantities.
For the metric (\ref{ssmetfinal}), the null basis or the
Newman-Penrose (NP) vectors are:
\beq \sigma_1 &= \ell &= \frac{1}{\sqrt{2}} \left[ \frac{1}{y}\frac{\partial}{\partial t} + \frac{1}{\gamma y' B} \frac{\partial}{\partial r}\right] \\
\sigma_2  &= n &=  \frac{1}{\sqrt{2}} \left[ \frac{1}{y}\frac{\partial}{\partial t} - \frac{1}{\gamma y' B} \frac{\partial}{\partial r}\right]\\
\sigma_3 &= m &=  \frac{1}{\sqrt{2}yY} \left[ \frac{\partial}{\partial \theta} - i \csc{\theta} \frac{\partial}{\partial \phi}\right]\\
\sigma_4 &=\bar{m} &=  \frac{1}{\sqrt{2}yY} \left[
\frac{\partial}{\partial \theta}+  i \csc{\theta}
\frac{\partial}{\partial \phi}\right]\eeq One notes that in a null
basis the only non-vanishing scalar products are $\ell \bullet n =
-1$ and $m\bullet\bar{m}=1$.

Then, the corresponding null 1-forms are:
\beq  \sigma^1 = \frac{1}{\sqrt{2}} [y dt + \gamma y' B dr ] \\
\sigma^2 = \frac{1}{\sqrt{2}} [ y dt - \gamma y' B dr ] \\
\sigma^3 = \frac{yY}{\sqrt{2}} [ d\theta + i \sin{\theta} d\phi ] \\
\sigma^4 = \frac{yY}{\sqrt{2}} [ d\theta - i \sin{\theta} d\phi ]
\eeq

The non-vanishing NP Spin coefficients are, then, obtained as:
\beq \alpha &=& - \beta = - \frac{\cot{\theta}}{2\sqrt{2}yY} \\
\epsilon &=& \frac{1}{2\sqrt{2}y}\left[ \frac{\dot{B}}{B} -
\frac{\dot{Y}}{Y} \right] \\
\mu &=& \frac{1}{\sqrt{2}y}\left[ \frac{\dot{Y}}{Y} - \frac{1}{\gamma B} \right] \\
\gamma &=& \frac{1}{2\sqrt{2}y}\left[ \frac{2}{\gamma B} - \frac{\dot{B}}{B}
- \frac{\dot{Y}}{Y} \right] \\
\rho &=& -\frac{1}{\sqrt{2}y}\left[ \frac{\dot{Y}}{Y} +
\frac{1}{\gamma B} \right] \eeq

Now, the non-vanishing components of the Weyl tensor for
(\ref{ssmetfinal}) are: \beq \label{weyl} C_{trtr} &=&
\frac{B^2\gamma^2(y')^2}{3}\,F(t)  \\
C_{t\theta t\theta} &=& -\,\frac{y^2Y^2}{6}\,F(t)  \\ C_{t\phi
t\phi} &=& \sin^2{\theta}\,C_{t\theta t\theta}
\\ C_{r\theta r\theta} &=&\frac{B^2\gamma^2Y^2(y')^2}{6}\, F(t) \\ C_{r\phi
r\phi} &=& \sin^2{\theta}\,C_{r\theta r\theta}
\\ C_{\theta\phi\theta\phi} &=& -\,\frac{y^2Y^4\sin^2{\theta}}{3}\,F(t)  \eeq where
\be F(t) = \frac{\ddot{Y}}{Y} -\frac{\dot{Y}^2}{Y^2}-\frac{1}{Y^2}
-\frac{\ddot{B}}{B}+\frac{\dot{B}\dot{Y}}{BY} \ee

As can be easily verified, the NP complex scalars  \beq \Psi_0
&\equiv -\,C_{pqrs}\ell^pm^q\ell^rm^s &= 0 \\ \Psi_1 &\equiv
-\,C_{pqrs}\ell^pn^q\ell^rm^s &= 0 \\\Psi_3 &\equiv
-\,C_{pqrs}\ell^pn^q\tilde{m}^rn^s &= 0 \\\Psi_4 &\equiv
-\,C_{pqrs}n^p\tilde{m}^qn^r\tilde{m}^s &= 0 \eeq and, hence, that
both the NP-vectors $\ell$ and $n$ are aligned along repeated
principal null directions of the Weyl tensor. The spacetime of
(\ref{ssmetfinal}) is therefore a Petrov-type D spacetime. 

It is well-known \cite{mtbh} that the shear-free geodesic
condition on both $\ell$ and $n$ in Petrov-type D spacetimes
ensures that the NP spin coefficients $\kappa$, $\sigma$,
$\lambda$, $\nu$ vanish as is evident from the NP spin
coefficients listed earlier.

\section{Concluding Remarks}
The black-hole spacetimes of General Relativity are all of
Petrov-type D \cite{mtbh}. It is, therefore, generally believed
that a Petrov-type D spacetime is required to describe the {\em
physically realizable\/} spherical gravitational collapse of
matter. That this is true for spherical symmetry is what is the
premise of the present paper.

It is well-known that Penrose \cite{penroseweyl} is led to the
Weyl hypothesis on the basis of thermodynamical considerations, in
particular, those related to the thermodynamic arrow of time. On
the basis of these considerations, we may consider the Weyl tensor
to be ``some" sort of measure of the {\em entropy\/} in the
spacetime at any given epoch.

Then, for non-singular and non-degenerate data in
(\ref{ssmetfinal}), the Weyl tensor of (\ref{ssmetfinal}) blows up
at the singular hyper-surface of (\ref{ssmetfinal}) but is
``vanishing" at the ``initial" hyper-surface since $\dot{Y} =
\dot{B} =0$ for the ``initial" hyper-surface \cite{prl1}.

This behavior of the Weyl tensor of (\ref{ssmetfinal}) is in
conformity with Penrose's Weyl curvature hypothesis
\cite{penroseweyl}. Thus, the spacetime of (\ref{ssmetfinal}) has
the ``right" kind of thermodynamic arrow of time in it.

As a separate remark, we note that, following the works of Ellis
and Sciama \cite{ellissciama}, there is an interpretation
\cite{tod} of Mach's principle, namely that there should be no
source-free contributions to the metric or that there should be no
source-free Weyl tensor for a Machian spacetime.

We, therefore, also note here that the vacuum is a degenerate case
for (\ref{ssmetfinal}). Then, the metric (\ref{ssmetfinal}) has no
source-free contributions and the spacetime of (\ref{ssmetfinal})
has no source-free Weyl tensor since the data is required to be
non-singular and non-degenerate for it. The spacetime of
(\ref{ssmetfinal}) is, then, Machian in this sense.

\acknowledgements{ We are grateful to Ravi Saraykar and Pradeep
Muktibodh for verifying some of the calculations. Some of the
reported calculations have been performed using the software {\tt
SHEEP} and SMW is indebted to Malcolm MacCallum for providing this
useful package to him.}

\end{document}